\documentclass[fleqn,twoside]{article}
\usepackage[headings]{espcrc2}

\readRCS
$Id: espcrc2.tex,v 1.2 2004/02/24 11:22:11 spepping Exp $
\usepackage{graphicx}
\usepackage[figuresright]{rotating}

\newcommand{\AmS}{{\protect\the\textfont2
  A\kern-.1667em\lower.5ex\hbox{M}\kern-.125emS}}

\title{Searches for New Physics at HERA}

\author{E. Sauvan\address[CPPM]{CPPM, IN2P3-CNRS et Universit\'e de la M\'editerran\'ee, 163 Av. de Luminy\\
 F-13288 Marseille, France}%
\thanks{On behalf of the H1 and ZEUS Collaborations.}
}
       
\runtitle{Searches for New Physics at HERA}
\runauthor{E. Sauvan}

\begin{document}

\begin{abstract}
The high energy programme of the HERA collider ended in March 2007.
During the whole HERA programme, a combined total integrated luminosity of $1$ fb$^{-1}$ was 
collected by the H1 and ZEUS experiments.
In this context, an overview of the most recent results of both experiments concerning searches
for new physics is presented. 
The topics covered are searches for contact interactions, leptoquarks 
and excited leptons, as well as studies of the isolated lepton and multi-lepton topologies, and
a general signature based search. 

\vspace{1pc}
\end{abstract}

\maketitle
\section{Introduction}

At HERA  electrons (or positrons) collide with protons at $\sqrt s \simeq 320$~GeV.
After two running periods, the high energy data taking ended in March 2007.
Over the whole HERA running the H1 and ZEUS experiments have each recorded $\sim 0.5$ fb$^{-1}$ of data, shared between $e^+p$ and $e^-p$ collision modes.
These high energy electron-proton interactions provide a testing ground for the Standard Model (SM) complementary to $e^+e^-$ and $p\bar{p}$ scattering, giving access to rare processes with cross sections below $1$~pb.
They are therefore used to pursue a rich variety of searches for new phenomena.

\section{Search for new phenomena in inclusive DIS}

\begin{figure}[htbp] 
  \begin{center}
    \includegraphics[width=.34\textwidth]{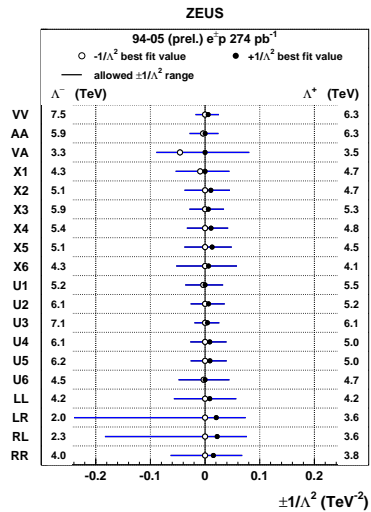}
  \end{center}
  \vspace*{-35pt}
  \caption{Limits at $95$\% C.L. on $\Lambda$ for various contact interaction scenario from recent ZEUS data.}
\label{fig:CI}  
\end{figure}

Neutral Current Deep Inelastic Scattering (NC DIS) is measured at HERA
for values of the photon virtuality $Q^2$ up to about $40000$ GeV$^2$.
At the highest $Q^2$, a good agreement between the data and the SM expectation, derived from the DGLAP evolution of parton density functions determined at lower $Q^2$,  
is observed. Stringent constraints on new physics can thus be set.
For example, a finite quark radius would
reduce the high $Q^2$ DIS cross section with respect to the SM predictions.
Using $1994$--$2005$ data with an integrated luminosity of $274$ pb$^{-1}$, the ZEUS Collaboration ruled out quark radii larger than $0.67 \times 10^{-18}$~m,
assuming that the electron is point-like~\cite{ZEUS_CI}.
New interactions between electrons and quarks with caracteristic mass scale in the TeV range may also modify the cross section at high $Q^2$ via virtual effects.
Many such interactions can be modelled as a four-fermion contact interaction. 
Currently, the ZEUS data constrain the effective scale $\Lambda$ of $eeqq$ contact interactions to be larger than $2$ to $7.5$ TeV, depending on the model, as presented in Figure~\ref{fig:CI}.

\section{Model dependent searches}

\subsection{Leptoquarks}

An intriguing characteristic of the Standard Model is the observed symmetry between the lepton and quark sectors.
This could be a possible indication of a new symmetry between the lepton and quark sectors, leading to ``lepto-quark'' interactions.
Leptoquarks (LQs) are new scalar or vector color-triplet bosons, carrying a fractional electromagnetic charge and both a baryon and a lepton number. Several types of LQs might exist, differing in their quantum numbers.
A classification of LQs has been proposed by Buchm\"uller, R\"uckl and Wyler (BRW)~\cite{brw_lq} under the assumptions that LQs have pure chiral couplings to SM fermions of a given family. The interaction of the LQ with a lepton-quark pair is of Yukawa or vector nature and is parametrised by a coupling $\lambda$.
%

\begin{figure}[htbp] 
  \begin{center}
    \includegraphics[width=.42\textwidth]{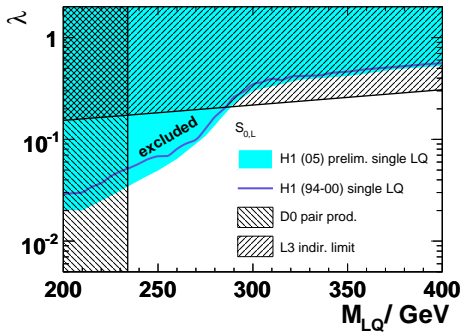}
  \end{center}
  \vspace*{-35pt}
  \caption{Exclusion limits at $95$\% C.L. on the coupling $\lambda$ as a function of the leptoquark mass for $S_{0,L}$ in the BRW model. The indirect limit from L3~\protect{\cite{Acciarri:2000uh}} and the direct D0 limits~\protect{\cite{Abazov:2004mk}} are also shown. For comparison, the published H1 limit on $S_{0,L}$ using HERA I data~\protect{\cite{Aktas:2005pr}} is also displayed.}
\label{fig:LQs}  
\end{figure} 

A new analysis has been performed by the H1 experiment using $92$~pb$^{-1}$ of data recorded in $2005$ in $e^-p$ collision mode~\cite{H1_LQ}. This large set of $e^-p$ data gives an increased the sensitivity to leptoquarks with fermion number $F=2$.
For leptoquark couplings of electromagnetic strength ($\lambda^2/4\pi = \alpha_{em}$), $F=2$ leptoquarks with masses up to $276$--$304$ GeV are now ruled out, depending on the LQ type.
As an example, the new constraint on $S_{0,L}$ leptoquarks are presented in Figure~\ref{fig:LQs} and compared to existing limits obtained at LEP by the L3 Collaboration~\cite{Acciarri:2000uh} or at the Tevatron by the D0 experiment~\cite{Abazov:2004mk}.

\subsection{Excited fermions}

The observed replication of three fermion families motivates the possibility of a new scale of matter yet unobserved.
An unambiguous signature for a new scale of matter would be the direct observation of excited states of fermions ($f^*$), via their decay into a gauge boson and a fermion. Effective models describing the interaction of excited fermions with standard matter have been proposed~\cite{Hagiwara:1985wt,Boudjema:1992em,Baur:1989kv}.
In the models~\cite{Hagiwara:1985wt,Boudjema:1992em} the interaction of an $f^*$ with a gauge boson is described by a magnetic coupling proportional to $1/\Lambda$ where $\Lambda$ is a new scale. Proportionality constants $f$, $f'$ and $f_s$ result in different couplings to $U(1)$, $SU(2)$ and $SU(3)$ gauge bosons.

\begin{figure}[htbp] 
  \begin{center}
    \includegraphics[width=.36\textwidth]{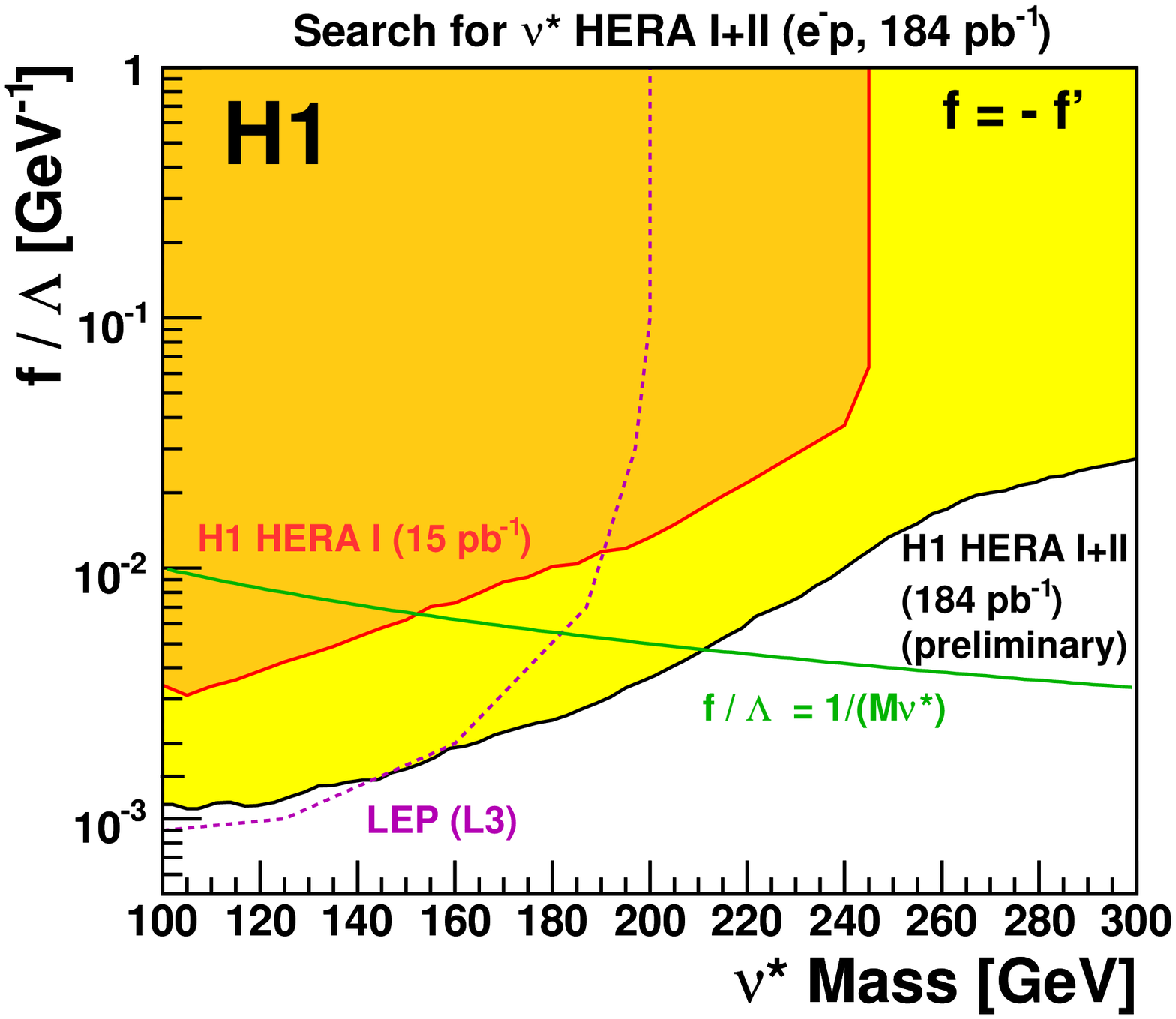}\put(-25,30){{\bf (a)}} \\
    \includegraphics[width=.36\textwidth]{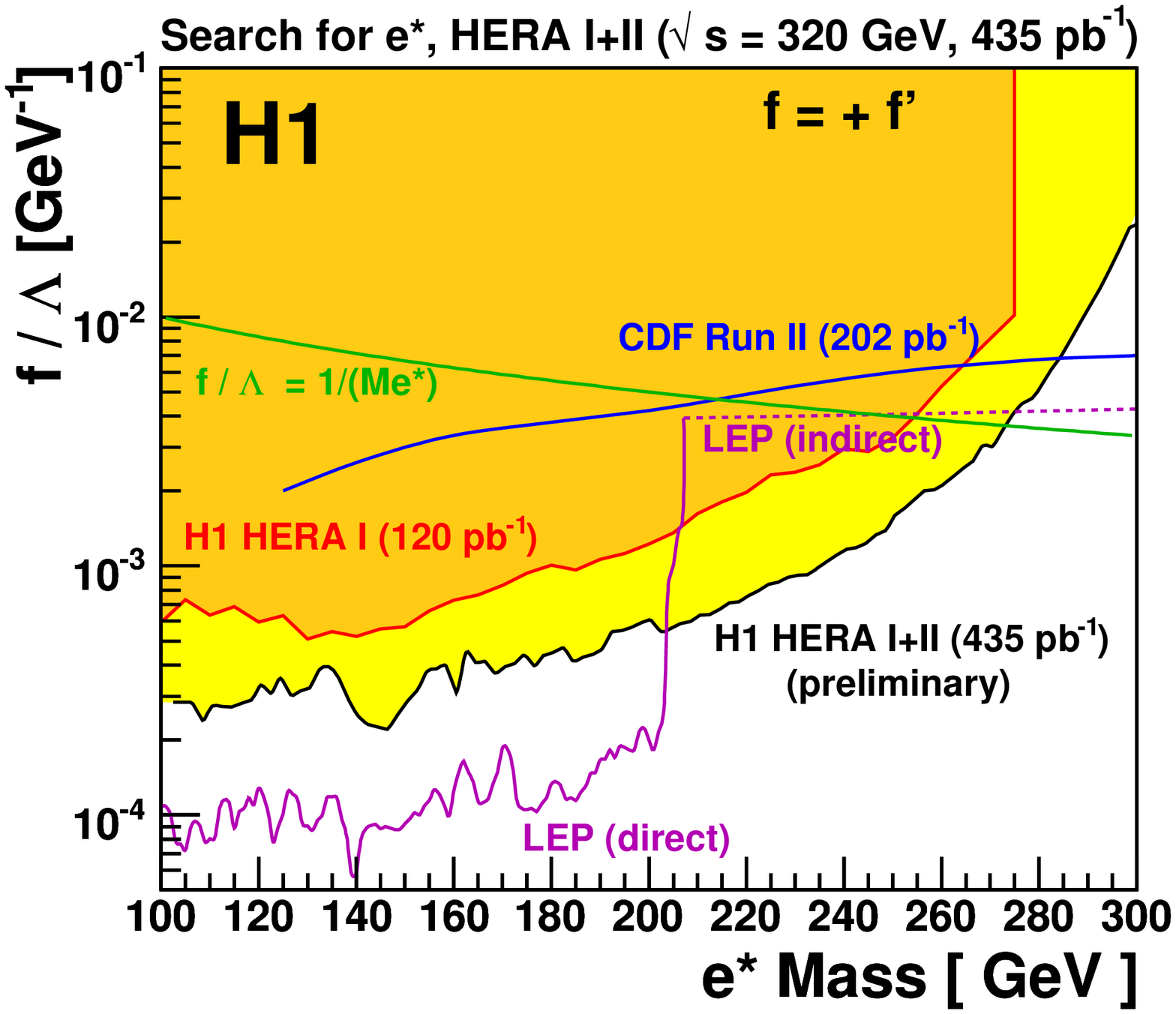}\put(-25,30){{\bf (b)}}
  \end{center}
  \vspace*{-35pt}
  \caption{Exclusion limits on the coupling $f/\Lambda$ at $95$\% C.L. as a function of the mass of the excited neutrino (a) and electron (b) with the assumptions $f = -f'$ and $f = +f'$, respectively. 
The new limits set by H1 are represented by the yellow area.
Values of the couplings above the curves are excluded.
}
\label{fig:exc_fermions}  
\end{figure} 

The H1 experiment has carried out searches for both excited neutrinos and electrons using all data recorded at $\sqrt s = 320$~GeV. The total luminosity analysed amount to up to $435$ pb$^{-1}$~\cite{H1_fstar}. The new bounds on the $\nu^*$ and $e^*$ masses obtained as a function of $f/\Lambda$ are presented in Figure~\ref{fig:exc_fermions}(a) and (b), under the assumptions $f = - f'$ and $f = + f'$, respectively. 
Assuming $f/\Lambda = 1/M_{\nu^*}$ and $f = - f'$, masses below $211$ GeV are ruled out for $\nu^*$. Excited electrons of mass below $273$ GeV are excluded at $95$\% C.L. if we assume $f/\Lambda = 1/M_{e^*}$ and $f = + f'$.
As observed in Figure~\ref{fig:exc_fermions}, the H1 analysis has probed new parameter space regions and limits set extend at high masses previous bounds reached at LEP and Tevatron colliders.

\begin{table*}
\begin{center}
\begin{tabular}{ccccc}
\hline
                 & Electron & Muon & Combined\\
$P_T^X > 25$ GeV & obs./exp. & obs./exp. & obs./exp. \\
\hline
~~H1~~ ~$e^-p$~~ $184$ pb$^{-1}$ & $3$ / $3.8 \pm 0.6$ & $0$ / $3.1 \pm 0.5$  & $3$ / $6.9 \pm 0.6$\\
ZEUS ~$e^-p$~~ $204$ pb$^{-1}$ & $5$ / $3.8 \pm 0.6$ & $2$ / $2.2 \pm 0.3$   & $7$ / $6.0 \pm 0.6$\\
\hline
~~H1~~ ~$e^+p$~~ $294$ pb$^{-1}$ & $11$ / $4.7 \pm 0.9$ & $10$ / $4.2 \pm 0.7$ & $21$ / $8.9 \pm 1.5$\\
ZEUS ~$e^+p$~~ $228$ pb$^{-1}$ & $1$ / $3.2 \pm 0.4$ & $3$ / $3.1 \pm 0.5$     & $4$ / $6.3 \pm 0.5$\\
\hline
\end{tabular}
\end{center}
\caption{Comparison of the number of isolated lepton (electron or muon) events observed for $P_T^X > 25$ GeV by H1 and ZEUS experiments with SM predictions.}
\label{tab:isol_lep_h1zeus}
\end{table*}

\section{Search for deviations from the SM in rare processes}

\subsection{Events with high $P_T$ isolated leptons}

\begin{figure}[htbp] 
  \begin{center}
    \includegraphics[width=.36\textwidth]{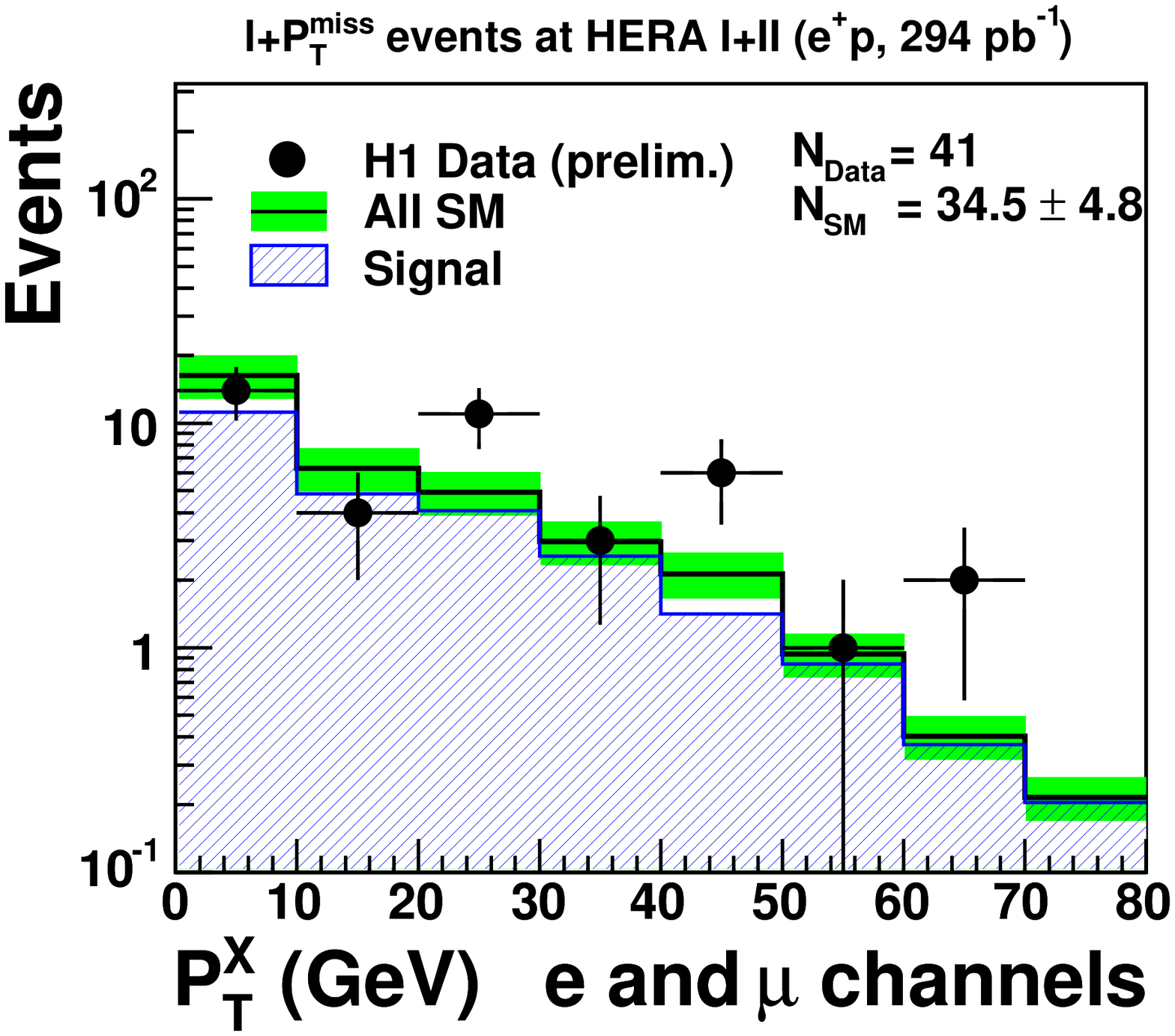}\put(-25,95){{\bf (a)}}\\
    \includegraphics[width=.36\textwidth]{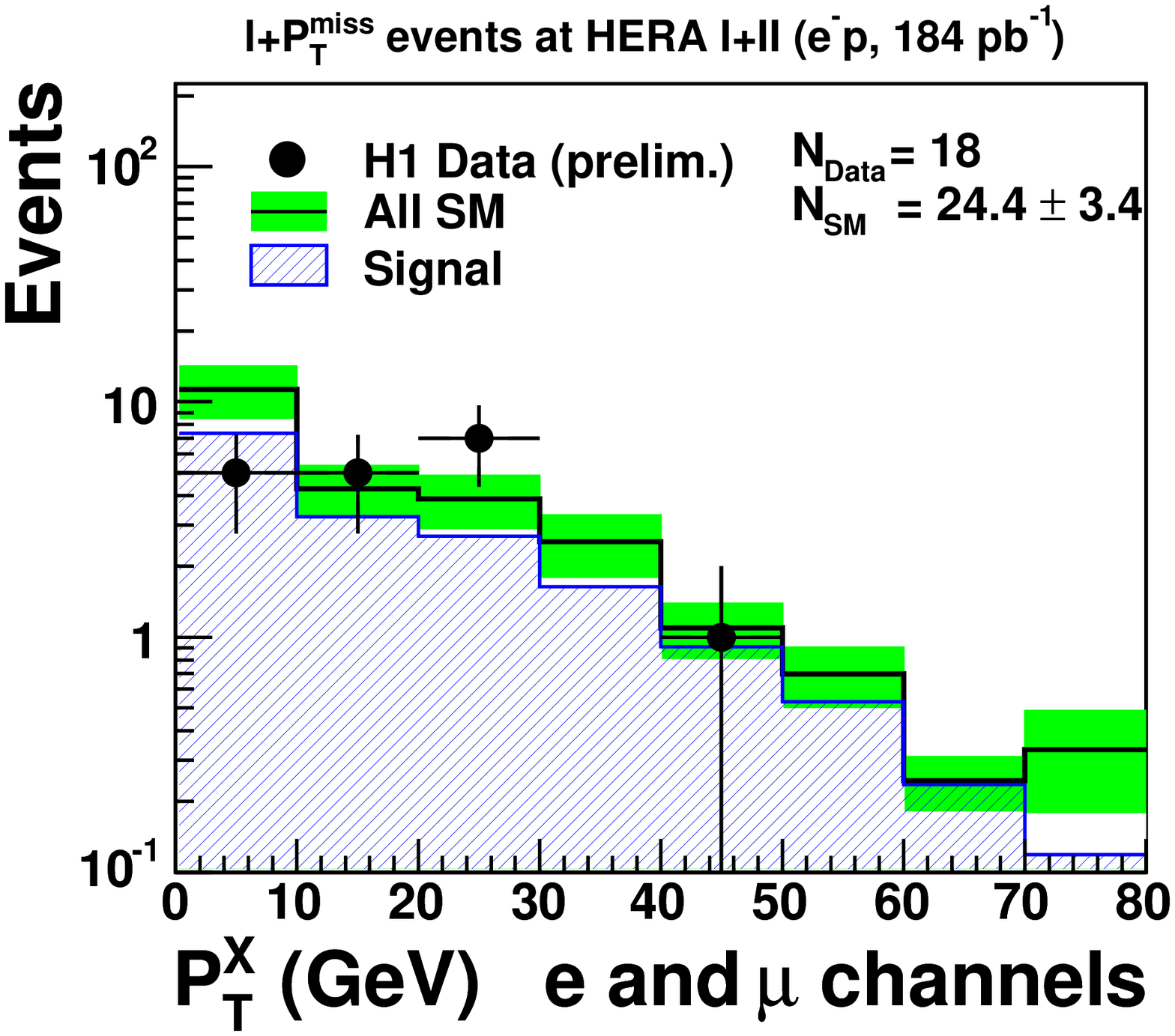}\put(-25,95){{\bf (b)}}
  \end{center}
  \vspace*{-35pt}
  \caption{Hadronic transverse momentum distribution of isolated lepton events observed by H1 in $e^+p$ (a) and $e^-p$ (b) data samples. The total SM expectation is represented by the open histograms and the contribution from $W$ production by the hatched histogram. }
\label{fig:isollep}  
\end{figure}

The production of a $W$ boson in $ep$ collisions at HERA has a cross-section of about $1$ pb. The leptonic decay of the $W$ leads to events with an isolated high transverse momentum lepton (electron, muon or tau) and missing total transverse momentum. Of particular interest are events with a hadronic system of large transverse momentum ($P_T^X$). An abnormally large rate of high $P_T^X$ events is observed by the H1 experiment~\cite{isollep_h1} in the electron an muon channels. In the analysis of all HERA I and HERA II data sets, which amounts to a total luminosity of $478$ pb$^{-1}$, $24$ events are observed at $P_T^X > 25$ GeV for a SM expectation of $15.8 \pm 2.5$. Amongst them only $3$ events are observed in $e^-p$ collisions, in agreement with the SM expectation, while $21$ events are observed in the $e^+p$ data for an expectation of $8.9 \pm 1.5$ (see Table~\ref{tab:isol_lep_h1zeus}). 
This difference in observations between the $e^+p$ and $e^-p$ data sets is exemplified in Figure \ref{fig:isollep} where the $P_T^X$ distributions of both data sets are displayed.

The ZEUS experiment has carried out a similar analysis using $432$ pb$^{-1}$ of $1996$--$2006$ data~\cite{ZEUS_isollep}. 
The results are also shown in Table~\ref{tab:isol_lep_h1zeus}. At $P_T^X > 25$~GeV the number of data events observed by ZEUS is in agreement with the SM expectation in both $e^+p$ and $e^-p$. 
A detailed comparison between efficiencies of the H1 and ZEUS detectors for the $W$ signal was performed. Both efficiencies are comparable in the central region, while H1 detection region extends to lower polar angle than ZEUS. Nevertheless most of the high $P_T^X$ events observed by H1 are lie within the range of the ZEUS acceptance.

The analysis of the tau decay channel is also performed by H1 on all HERA data with a total luminosity of $471$~pb$^{-1}$. In this channel, the separation of the $W$ signal from other SM processes is more difficult and  the purity and efficiency are lower than for the $e$ and $\mu$ channels. In total $20$ data events are observed compared to a SM expectation of $19.5 \pm 3.2$. One of the data events has $P_T^X$ above $25$ GeV, compared to a SM expectation of $0.99 \pm 0.13$.
%
%

\subsection{Multi-lepton events}

Multi-lepton production has also been studied at HERA. Here the main production mechanism is photon-photon collisions. All event topologies with high $P_T$ electrons and muons have been investigated by the H1 experiment using a total luminosity of $459$ pb$^{-1}$~\cite{H1_mlep}. 
The measured yields of di-lepton and tri-lepton events are in good agreement with the SM prediction, except in the tail of the distribution of the scalar sum of $P_T$ of the leptons ($\sum P_T$). In $e^+p$ collisions, $4$ data events are observed with $\sum P_T > 100$ GeV compared to a SM prediction of $1.2 \pm 0.2$. No such events are observed in $e^-p$ collisions for a similar SM expectation of $ 0.8 \pm 0.2$.

The analysis of $ee$ and $eee$ topologies is also carried out by ZEUS using $446$ pb$^{-1}$ of data~\cite{ZEUS_mlep}. In this analysis, the SM contribution at high invariant mass in the di-electron channel is dominated by Compton background due to a more difficult $e / \gamma$ separation in the ZEUS detector. Nevertheless $5$ and $1$ events with a high invariant mass above $100$ GeV are observed in $ee$ and $eee$ channels, respectively. This observation is in good agreement with the corresponding SM expectations of $3.4 ^{+0.6}_{-0.3}$ and $1.1 ^{+0.5}_{-0.1}$, respectively.

\subsection{A general search for new phenomena}

\begin{figure}
  \begin{center}
    \includegraphics[width=.5\textwidth,angle=-90]{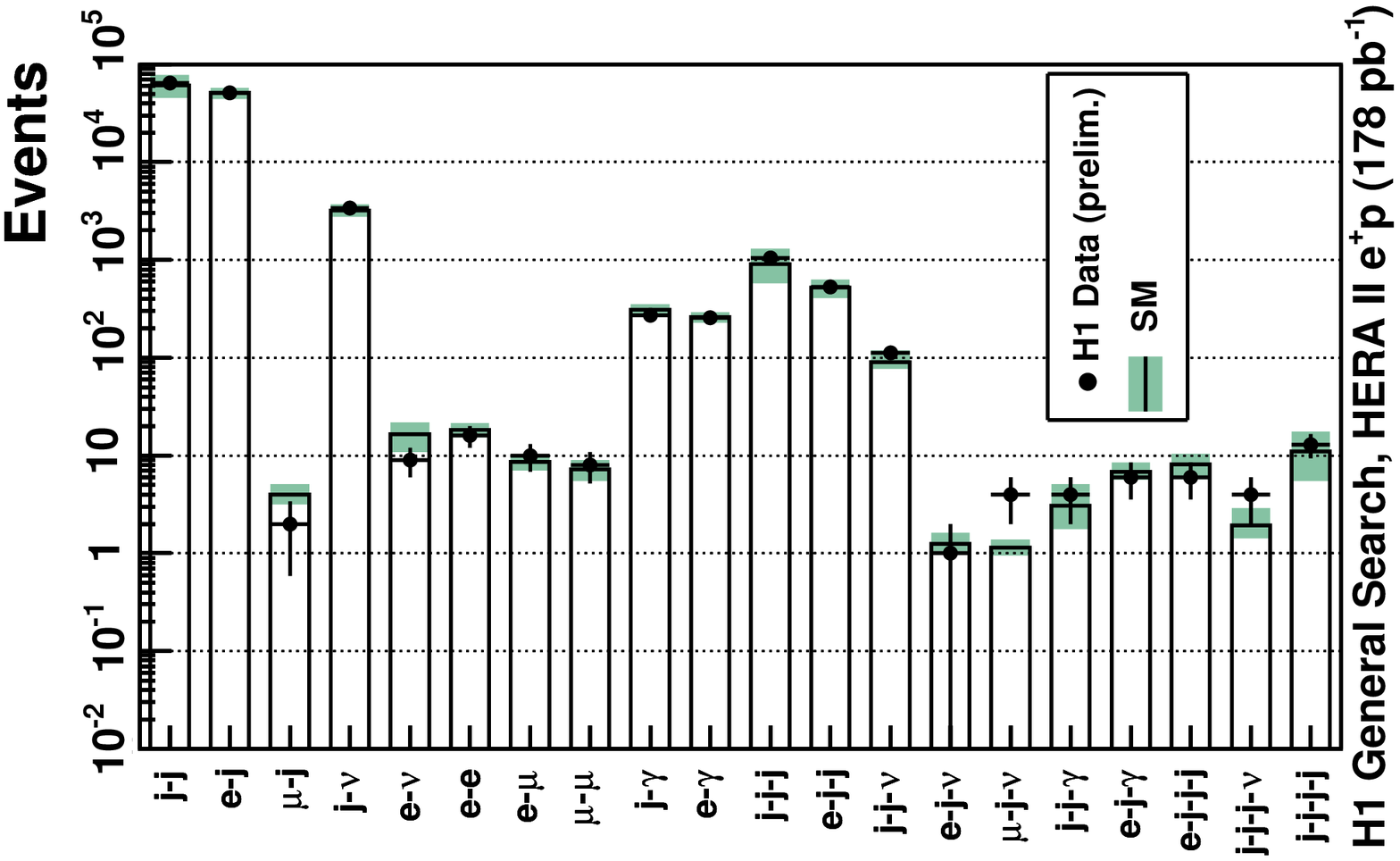}\put(-30,-207){{\bf (a)}}\\
    \includegraphics[width=.5\textwidth,angle=-90]{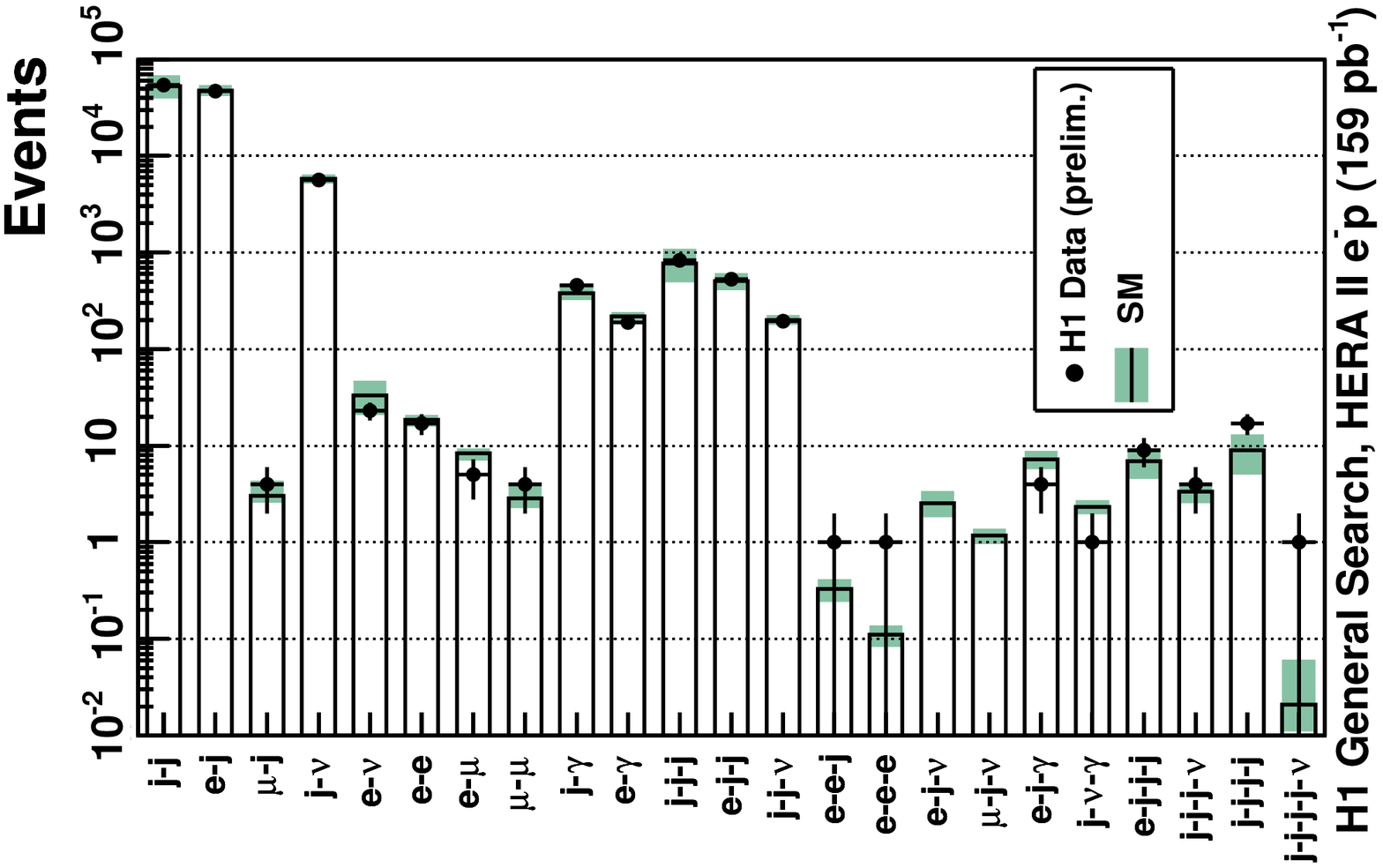}\put(-30,-207){{\bf (b)}}
  \end{center}
  \vspace*{-25pt}
  \caption{The data and the SM expectation in event classes investigated by the H1 general search. All channels with observed data events or a SM expectation greater than one event are displayed. The results are presented separately for $e^+p$ (a) and $e^-p$ (b) collision modes.}
\label{fig:GS}  
\end{figure}

In a more general way, a broad range signature based search has been developed by the H1 Collaboration~\cite{Aktas:2004pz}. The same analysis is now applied to all HERA II data~\cite{H1_GS} and all final states containing at least two objects ($e$, $\mu$, $j$, $\gamma$, $\nu$) with 
$P_T >$~$20$~GeV in the polar angle range  $10^\circ < \theta < 140^\circ$ have been investigated. The observed and predicted yields in each channel are presented in Figure~\ref{fig:GS}(a) and (b) for $e^+p$ and $e^-p$ collisions, respectively. The good agreement observed between data and SM prediction demonstrates the good understanding of the detector and of the contributions of the SM backgrounds. 
A systematic scan of $M_{all}$ and $\sum P_T$ distributions in each channel has been performed to look for regions of largest deviations to the SM. A statistical analysis is then used to quantify the significance of observed deviations. The largest deviation is observed in $e^+p$ data in the $\mu$-$j$-$\nu$ channel which corresponds to the topology of isolated lepton events.

\section{Conclusions}

The most recent results of searches for new physics performed at the HERA $ep$ collider have been presented.
Most of the analyses fully exploit the complete data sample available which amounts to $\sim 0.5$ fb$^{-1}$ per experiment.
HERA appears to be very well suited to search for new phenomena in specific cases, complementary to stringent bounds set at LEP and the Tevatron.
Nevertheless, no convincing evidence for the existence of new phenomena has been observed so far.
Among all event topologies investigated, the largest deviation to the SM expectation is observed by the H1 experiment for isolated lepton events in $e^+p$ collisions only. After having analysed all data recorded by H1, this deviation corresponds to a $3$ $\sigma$ excess of atypical $W$-like events.

\end{document}